\documentclass[aps,prd,showpacs,nofootinbib,preprintnumbers,floatfix,twocolumn]{revtex4}
\usepackage{bm}
\usepackage{latexsym}
\usepackage{color}
\usepackage{dcolumn}
\usepackage{amsfonts,amssymb,amsmath}
\usepackage{graphicx,epsfig}
\usepackage{psfrag}
\usepackage{subfigure}
\def\beq{\begin{equation}}
\def\eeq{\end{equation}}
\def\br{\begin{eqnarray}}
\def\er{\end{eqnarray}}
\def\benu{\begin{enumerate}}
\def\eenu{\end{enumerate}}

\def\l{\left}
\def\r{\right}
\def\d{{\rm d}}

\def\f{\frac}
\begin{document}
%%%%%%%%%%%%%%%%%%%%%%%%%%%%%%%%%%%%%%%%%%%%%%%%%%%%%%%%%%%%%%%%%%%%%%%%%%%%%%%
\title{Oscillations in the inflaton potential:~Complete numerical treatment\\ 
and comparison with the recent and forthcoming CMB datasets}
%%%%%%%%%%%%%%%%%%%%%%%%%%%%%%%%%%%%%%%%%%%%%%%%%%%%%%%%%%%%%%%%%%%%%%%%%%%%%%%
\author{Moumita Aich$^{1}$\footnote{Current address:~Astrophysics 
and Cosmology Research Unit, School of Mathematical Sciences,
University of Kwazulu-Natal, Durban, South Africa.
E-mail:~moumita@iucaa.ernet.in},
Dhiraj Kumar Hazra$^{2}$\footnote{E-mail:~dhiraj@hri.res.in}, 
L.~Sriramkumar$^{2}$\footnote{Current address:~Department of Physics,
Indian Institute of Technology Madras, Chennai~600036, India. 
E-mail:~sriram@physics.iitm.ac.in}
and Tarun Souradeep$^{1}$\footnote{E-mail:~tarun@iucaa.ernet.in}}
%%%%%%%%%%%%%%%%%%%%%%%%%%%%%%%%%%%%%%%%%%%%%%%%%%%%%%%%%%%%%%%%%%%%%%%%%%%%%%%
\affiliation{$^{1}$IUCAA, Post Bag 4, Ganeshkhind, Pune 411007, India\\
$^{2}$Harish-Chandra Research Institute, Chhatnag Road,
Jhunsi, Allahabad~211019, India}
\date{\today}
\begin{abstract}
Amongst the multitude of inflationary models currently available, models 
that lead to features in the primordial scalar spectrum are drawing 
increasing attention, since certain features have been found to provide a 
better fit to the CMB data than the conventional, nearly scale invariant,
primordial spectrum. 
In this work, we carry out a complete numerical analysis of two models that 
lead to oscillations {\it over all scales}\/ in the scalar power spectrum.
We consider the model described by a quadratic potential which is superposed 
by a sinusoidal modulation and the recently popular axion monodromy model. 
Since the oscillations continue even on to arc minute scales, in addition 
to the WMAP data, we also compare the models with the small scale data 
from ACT.
Though, both the models, broadly, result in oscillations in the spectrum, 
interestingly, we find that, while the monodromy model leads to a 
considerably better fit to the data in comparison to the standard power 
law spectrum, the quadratic potential superposed with a sinusoidal 
modulation does not improve the fit to a similar extent.
We also carry out forecasting of the parameters using simulated Planck 
data for both the models. 
We show that the Planck mock data performs better in constraining the 
model parameters as compared to the presently available CMB datasets. 
\end{abstract}
\maketitle

%%%%%%%%%%%%%%%%%%%%%%%%%%%%%%%%%%%%%%%%%%%%%%%%%%%%%%%%%%%%%%%%%%%%%%%%%%%%%%%

\section{Local versus non-local features in the primordial spectrum}
 
Inflation has proven to be the most efficient mechanism to overcome
difficulties such as the horizon and the flatness problems that plague 
the standard, hot, big bang cosmological model.
Importantly, in addition to resolving such issues, inflation successfully
generates primordial fluctuations which seed the formation of structures. 
Over the years, numerous models have been proposed that lead to a sufficient 
duration of inflation and also produce perturbations of suitable amplitude 
and shape that are consistent with the observations of the anisotropies
in the Cosmic Microwave Background (CMB)~\cite{wmap-5,wmap-7,act-2010} 
as well as other observational bounds. 
Currently, many of the models that lead to slow roll inflation and, 
therefore, to a nearly scale invariant primordial spectrum, seem to 
perform equally well against the available data~\cite{martin-2011}.
The hope remains that further CMB data from the present and forthcoming 
missions such as Planck~\cite{planck} and the Cosmic Origins 
Explorer~\cite{core} may help us discriminate better between 
the various models.

\par

Although, a nearly scale invariant power spectrum predicted by slow roll
inflation, along with the background $\Lambda$CDM model, matches the angular 
spectrum from  the CMB observations quite well, there exist a few outliers 
(notably, near the multipole moments of $\ell=2$, $22$ and $40$) in the 
Wilkinson Microwave Anisotropy Probe (WMAP) data~\cite{wmap-5,wmap-7}. 
Interestingly, model independent reconstruction of the primordial spectrum 
from the observed pattern of the CMB anisotropies seem to suggest the
possible presence of specific features in the spectrum (see Refs.~\cite{rc}; 
for a different view, see Refs.~\cite{rc-wf}).
Moreover, it has been found that certain localized features actually lead to 
a better fit to the data than the conventional power law spectrum (in this
context, see, for instance, 
Refs.~\cite{wmap-1,quadrupole,pi,l2240,general}).
And, it should be noted here that generating such features require either
one or more periods of deviation from slow roll 
inflation~\cite{starobinsky-1992,dvorkin-2010,hu-2011} or modifications
to the initial conditions on the perturbations~\cite{quadrupole-sic}.

\par

Apart from localized features, it is interesting to examine whether the CMB 
data also point to non-local features---i.e. certain characteristic and
repeated behavior that extend over a wide range of scales---in the primordial 
spectrum.
A quick glance at the unbinned CMB data seems to suggest that, after all,
such an eventuality need not altogether be surprising.
In fact, earlier investigations on possible Planck scale modifications to
the primordial spectrum have indicated that continuing oscillations in the
power spectrum can lead to a substantial improvement in the fit at the
cost of two or three additional parameters (see Refs.~\cite{pso-1,pso-2}; in 
this context, also see Ref.~\cite{ops}).
In this work, we shall investigate two inflationary models involving the
canonical scalar field that lead to similar oscillations over all scales 
in the curvature perturbation spectrum.
We shall consider a model described by the conventional quadratic potential,
but superposed by a sinusoidal modulation~(see Ref.~\cite{pahud-2009};
also see Ref.~\cite{chen-2008}), and the presently popular axion monodromy 
model (see, for example, Refs.~\cite{flauger-2010-2011}). 
It should be mentioned here that these models have been compared with the WMAP 
data recently~\cite{pahud-2009,flauger-2010-2011,kobayashi-2011}.
However, all the earlier analyses had resorted to evaluating the scalar 
power spectrum in the slow roll approximation. 
In contrast, we shall compute the scalar power spectrum {\it exactly}\/ 
using a highly accurate numerical code.
And, we shall evaluate the tensor spectrum too accurately and include 
it in our analysis.
Moreover, as the oscillations in the inflationary scalar power spectrum 
continue even over smaller scales, in addition to the WMAP seven year 
data~\cite{wmap-7}, we shall compare the models with the small scale data 
from the Atacama Cosmology Telescope (ACT)~\cite{act-2010}.
We shall also arrive at the constraints on the model parameters using 
simulated Planck data. 
While both the models that we consider lead to oscillations in the 
spectrum, we find that the monodromy model results in a superior 
fit to the data.
Further, as we shall see, the Planck mock data leads to better 
constraints on the model parameters than the currently available 
CMB datasets.  

\par

This paper is organized as follows.
In the following section, we shall briefly describe the models that we
shall consider and the methodology that we shall adopt to compare the 
models with the data.
In the subsequent two sections, we shall present the results of our 
analysis and examine whether Planck will be able to constrain the
models better, respectively.
We shall conclude with a brief summary and discussion in the final 
section.

\par

Note that we shall work in units such that $\hbar= c =(8\, \pi\, G)=1$.
Moreover, we shall assume the background cosmological model to be the 
standard, spatially flat, $\Lambda$CDM model.

%%%%%%%%%%%%%%%%%%%%%%%%%%%%%%%%%%%%%%%%%%%%%%%%%%%%%%%%%%%%%%%%%%%%%%%%%%%%%%%

\section{Models and methodology}

In this section, we shall briefly describe the models that we shall work 
with and the methodology we shall adopt to compare the models with the
data. 

%%%%%%%%%%%%%%%%%%%%%%%%%%%%%%%%%%%%%%%%%%%%%%%%%%%%%%%%%%%%%%%%%%%%%%%%%%%%%%%

\subsection{The models}

As we mentioned, we shall consider two models, the first of which 
is the chaotic inflationary that is modulated by sinusoidal 
oscillations~\cite{pahud-2009,chen-2008}. 
The model is described by the potential 
\beq
V(\phi) = \frac{1}{2}\,m^2\,\phi^2  \l[1 + \alpha\,
\sin\l(\frac{\phi}{\beta}+\delta\r) \r],\label{eq:qpsin}
\eeq
where, evidently, $\phi$ denotes the canonical scalar field, $m$ is the 
parameter that characterizes the original quadratic potential, while the 
parameters~$\alpha$ and $\beta$ describe the amplitude and the frequency 
of the superimposed oscillations.
We have also included the parameter~$\delta$, which shifts the 
oscillations within one period, in our analysis. 
The second model that we shall consider is the axion monodromy model
which is motivated by string 
theory~\cite{flauger-2010-2011,kobayashi-2011,hannestad-2010,hazra-2012}.
The inflaton potential in such a case is given by
\beq
V(\phi)
=\lambda\, \l[\phi+\alpha\,\cos\l(\frac{\phi}{\beta}+\delta\r)\r].
\label{eq:amm}
\eeq
Note that, while the amplitude of the oscillation is fixed in the 
axion monodromy model, in the chaotic model described by the 
potential~(\ref{eq:qpsin}), the amplitude depends quadratically
on the field.
The inflaton oscillates as it rolls down these potentials, and these
oscillations continue all the way until the end of inflation.
This behavior leads to small oscillations in the slow roll parameters,
which in turn results in continuing oscillations in the primordial 
scalar power spectrum.
Our goal is to examine the extent to which such oscillations are admitted 
by the CMB data.

\par 

We shall compare the performance of the above two inflationary models with 
the conventional, power law, primordial spectrum.
Recall that, the power law, scalar and tensor spectra are usually written 
as (see, for example, Refs.~\cite{bassett-2006,sriram-2009})
\beq
{\cal P}_{_{\rm S}}(k)=A_{_{\rm S}} \l(\f{k}{k_{0}}\r)^{n_{_{\rm S}}-1}
\quad{\rm and}\quad
{\cal P}_{_{\rm T}}(k)=A_{_{\rm T}} \l(\f{k}{k_{0}}\r)^{n_{_{\rm T}}},
\label{eq:ps-pl}
\eeq
where the quantities $A_{_{\rm S}}$ and $A_{_{\rm T}}$ denote the amplitude 
of the scalar and tensor spectra, while $n_{_{\rm S}}$ and $n_{_{\rm T}}$
denote the corresponding spectral indices.
The quantity $k_{0}$ is the pivot scale at which the amplitudes of the 
power spectra are quoted.
Given the scalar and tensor spectra, the tensor-to-scalar ratio $r$ is defined 
as the ratio of the latter to the former and, when comparing the power law case 
with the observations, it is the quantity~$r$ that is usually considered in lieu
of the tensor amplitude $A_{_{\rm T}}$.
Also, when considering the power law spectra, as is often done, we shall 
assume the slow roll consistency condition (viz. that $r=-8\; n_{_{\rm T}}$), 
so that the power law case is essentially described by the three parameters 
$A_{_{\rm S}}$, $n_{_{\rm S}}$ and~$r$.

%%%%%%%%%%%%%%%%%%%%%%%%%%%%%%%%%%%%%%%%%%%%%%%%%%%%%%%%%%%%%%%%%%%%%%%%%%%%%%%

\subsection{Evaluation of the background and the perturbations}
\label{subsec:ebp}

We shall now outline the methods that we adopt to evolve the equations 
governing the background and the perturbations, and eventually 
evaluate the inflationary scalar and the tensor perturbation spectra.

\par 

Recall that, in a Friedmann universe, a canonical scalar field that 
is described by the potential $V(\phi)$ satisfies the following 
equation of motion:
\beq
\ddot{\phi}+3\,H\,\dot{\phi}+V_{\phi}=0,
\eeq
where $V_{\phi}=\d V/\d \phi$, $H$ is the Hubble parameter, while, 
as usual, the overdots denote differentiation with respect to the
cosmic time coordinate.
We solve the above differential equation exactly using the standard 
fourth order Runge-Kutta method, with e-folds as the independent 
variable. 
In the case of the chaotic inflationary model with sinusoidal modulations,
we choose the initial value of the field to be $\phi_{\rm i}\simeq 16$, while 
in the monodromy model, we set $\phi_{\rm i}\simeq 12$.
We then make use of the governing equations, considered under the slow roll 
approximation, to determine the initial velocities of the field.
These initial conditions allow sufficient number of e-foldings (say, about
$60$--$70$) before inflation ends near the bottom of the potentials. 
Further, following the convention, we shall choose the initial value of the
scale factor to be such that the pivot scale $k_0=0.05\; {\rm Mpc}^{-1}$ leaves 
the Hubble radius at $50$ e-folds before the end of inflation~\cite{l2240}.

\par

In the spatially flat Friedmann universe of our interest, the Fourier
modes of the curvature perturbation~$R$ and the tensor perturbation~${\tt h}$
are described by the following equations~\cite{bassett-2006,sriram-2009}:
\beq
R_{k}''+2\, \frac{z'}{z}\, R_{k}'+k^{2}\, R_{k} = 0\quad{\rm and}\quad
{\tt h}_{k}''+2\, \frac{a'}{a}\, {\tt h}_{k}'+k^{2}\, {\tt h}_{k} = 0,
\eeq
where the overprimes denote differentiation with respect to the conformal 
time coordinate and $z=a\,{\dot \phi}/H$, with $a$ being the scale factor.
We impose the standard Bunch-Davies initial conditions on the perturbations 
when the modes are well inside the Hubble radius, and evolve them using a 
Bulirsch-Stoer algorithm with an adaptive step size control routine~\cite{nr}.
In simpler and smoother inflationary potentials, the initial conditions on 
the modes are usually imposed when, say, $k/(a\, H) \simeq 100$.
In contrast, oscillatory potentials of our interest here can exhibit
certain resonant behavior and, in order to capture this behavior,
depending on the values of the potential parameters, it can become 
necessary to integrate from deeper inside the Hubble radius (in 
this context, see, for instance, Refs.~\cite{flauger-2010-2011}).
(It is interesting to note here that the resonance also leads 
to rather high levels of non-Gaussianities in these 
models~\cite{flauger-2010-2011,hannestad-2010,hazra-2012}.)
We impose the initial conditions on the modes when $k/(a\, H) \simeq 250$,
which we find to be suitable for the range of parameters of the 
potentials that we work with.
We evaluate the scalar and the tensor perturbation spectra, viz.
\beq
{\cal P}_{_{\rm S}}(k)=\f{k^2}{2\, \pi^{2}}\; \vert R_{k}\vert^{2}
\quad{\rm and}\quad
{\cal P}_{_{\rm T}}(k)=8\, \f{k^2}{2\, \pi^{2}}\; \vert {\tt h}_{k}\vert^{2},
\eeq
at super-Hubble scales, when the amplitude of the curvature and the tensor 
perturbations have frozen in [typically, when $k/(a\,H) \simeq 10^{-5}$].

%%%%%%%%%%%%%%%%%%%%%%%%%%%%%%%%%%%%%%%%%%%%%%%%%%%%%%%%%%%%%%%%%%%%%%%%%%%%%%%

\subsection{Priors}

As we mentioned, we shall assume the background cosmological model to be
the standard, spatially flat, $\Lambda$CDM model.
The model can be characterized by the following four 
parameters:~$\Omega_{\rm b}\, h^2$, $\Omega_{\rm c}\, h^2$,  $\theta$ 
and $\tau$.
The first two represent the baryon and the CDM densities (with $h$ being 
related to the Hubble parameter), while the last two denote the ratio of 
the sound horizon to the angular diameter distance at decoupling and the 
optical depth to reionization, respectively.
In Tab.~\ref{tab:priors-bp} below, we have listed the priors that we work
with on these four parameters.
%%%%%%%%%%%%%%%%%%%%%%%%%%%%%%%%%%%%%%%%%%%%%%%%%%%%%%%%%%%%%%%%%%%%%%%%%%%%%%%
\begin{table}[!htb]
\begin{center}
\begin{tabular}{|c|c|c|}
\hline
Background parameter & Lower limit & Upper limit \\ 
\hline
$\Omega_{\rm b}\, h^2$ & $0.005$ & $0.1$\\
\hline
$\Omega_{\rm c}\, h^2$ & $0.01$ & $0.99$\\
\hline
$\theta$ & $0.5$ & $10.0$\\
\hline
$\tau$ & $0.01$  & $0.8$\\
\hline
\end{tabular}
\caption{\label{tab:priors-bp}The priors on the four parameters that 
describe the background, spatially flat, $\Lambda$CDM model. 
We keep the same priors on the background parameters for all the models 
and datasets that we consider.}
\end{center}
\end{table}
%%%%%%%%%%%%%%%%%%%%%%%%%%%%%%%%%%%%%%%%%%%%%%%%%%%%%%%%%%%%%%%%%%%%%%%%%%%%%%%

\par

As we had discussed earlier, we shall include the tensor perturbations 
in our analysis. 
When the slow roll consistency condition is imposed, the power law spectra 
are completely described by the scalar amplitude $A_{_{\rm S}}$, the scalar 
spectral index $n_{_{\rm S}}$ and the tensor-to-scalar ratio $r$.
It is worth noting here that, in the inflationary models, the parameters
that describe the potential determine the scalar {\it as well as}\/ the
tensor spectra entirely.

\par
 
It is clear that, in the absence of the oscillatory terms in the 
potential, the two inflationary models of our interest will lead 
to nearly scale invariant spectra.
Therefore, the primary parameter that describes the two models, 
viz. $m$ in the chaotic inflationary model and $\lambda$ in the case 
of the axion monodromy model, are essentially determined by COBE 
normalization.
In the absence of oscillations in the potential, we find that the 
best fit chaotic model leads to a power law spectrum with a scalar 
spectral index of about $0.96$, while the monodromy model corresponds 
to $n_{_{\rm S}} \simeq 0.97$.
Also, as one would have anticipated, both of them perform almost equally 
well against the data.
However, when the oscillations in the potential are taken into account,
they induce modulations in the slow roll parameters, which in turn lead
to oscillations in the scalar power spectrum.
As we shall see, when the oscillations are included, the monodromy model 
performs better against the data than the chaotic inflation model.
 
\par

We have chosen the priors on the two inflationary models such that the
amplitude of the resulting scalar spectra remain close to the COBE value, 
lead to the desired spectral index, and result in a certain minimum 
duration of inflation.
The choice of priors have also been guided by the results from the 
earlier analysis~\cite{pahud-2009,flauger-2010-2011,kobayashi-2011}, 
and they allow us to capture the resonance that can arise in these 
models. 
We have listed the priors that we have worked with on the inflationary 
models in Tab.~\ref{tab:priors-ip}.
%%%%%%%%%%%%%%%%%%%%%%%%%%%%%%%%%%%%%%%%%%%%%%%%%%%%%%%%%%%%%%%%%%%%%%%%%%%%%%%
\begin{table}[!htb]
\begin{center}
\begin{tabular}{|c|c|c|c|}
\hline
Model & Parameter & Lower limit & Upper limit \\
\hline
& ${\rm ln}\, \l[10^{10}\, A_{_{\rm S}}\r]$  & $2.7$   & $4.0$\\
\cline{2-4}
Power law case & $n_{_{\rm S}}$ & $0.5$ & $1.5$\\
\cline{2-4}
& $r$ & $0.0$ & $1.0$\\
\hline
& ${\rm ln}\,  \l[10^{10}\, m^{2}\r]$ & $-0.77$ & $-0.58$   \\
\cline{2-4}
Chaotic model & $\alpha$ & $0$ & $2\times10^{-3}$\\
\cline{2-4}
with sinusoidal & $\beta$ & $2\times10^{-2}$ & $1$\\
\cline{2-4}
modulation & $\delta$ &$-\pi$ & $\pi$\\
\hline
& ${\rm ln}\,  \l[10^{10}\,\lambda\r]$ & $0.7$ &$1.25$   \\
\cline{2-4}
Axion monodromy  & $\alpha$ & $0$ & $2\times10^{-4}$\\
\cline{2-4}
model & $\beta$ & $3 \times 10^{-4}$ & $1\times10^{-3}$ \\
\cline{2-4}
& $\delta$ & $-\pi$ & $\pi$\\
\hline
\end{tabular}
\caption{\label{tab:priors-ip}The priors on the three parameters that 
describe the primordial spectra in the power law case, and the parameters
that describe the two inflationary potentials of our interest.
We work with the same priors when comparing the models with the WMAP as 
well as the ACT data.}
\end{center}
\end{table}

%%%%%%%%%%%%%%%%%%%%%%%%%%%%%%%%%%%%%%%%%%%%%%%%%%%%%%%%%%%%%%%%%%%%%%%%%%%%%%%

\subsection{Comparison with the recent CMB observations}

To compare our models with the recent CMB observations, we perform the 
by-now common practice of the Markov Chain Monte Carlo sampling of the 
parameter space using the publicly available CosmoMC 
package~\cite{cosmomc,lewis-2002}.
The CosmoMC code in turn utilizes the Boltzmann code CAMB~\cite{camb,lewis-2000} 
to arrive at the CMB angular power spectrum from given primordial scalar 
and tensor spectra. 
We evaluate the inflationary scalar as well as tensor spectra 
using an accurate and efficient numerical code (as outlined in 
Subsec.~\ref{subsec:ebp}) and feed these primordial spectra into
CAMB to obtain the corresponding CMB angular power spectra.
We should stress here that we actually evolve {\it all the modes}\/ that 
are required by CAMB from the sub to the super-Hubble scales to obtain the 
perturbation spectra, rather than evolve for a smaller set of modes and 
interpolate to arrive at the complete spectrum.
This becomes imperative in the models of our interest which (as one
would expect, and as we shall illustrate below) contain fine features 
in the scalar power spectrum.
It should be pointed out here that, while the chaotic model leads to 
a tensor-to-scalar ratio of $0.16$, the monodromy model results in 
$r\simeq  0.06$. 
Though these tensor amplitudes are rather small to make any significant
changes to the results, we have developed the code to evaluate the
inflationary power spectra with future datasets (such as, say, Planck) 
in mind, and hence we nevertheless take the tensors into account exactly. 

\par

For our analysis, we consider the WMAP seven year data and the small 
scale data from ACT~\cite{act-2010}.
We have worked with the May~2010 versions of the CosmoMC and CAMB 
codes~\cite{cosmomc,lewis-2002,camb,lewis-2000}, and we have made use 
of the WMAP (version v4p1) and the ACT likelihoods while comparing with 
the corresponding data~\cite{lambda}. 
While ACT has observed CMB at the frequencies of $148\; {\rm GHz}$ as 
well as $218\; {\rm GHz}$, we shall only consider the $148\; {\rm GHz}$ 
data. 
Moreover, though the ACT data spans over a wide range of multipoles
($500 \lesssim \ell \lesssim 10000$), for the sake of numerical 
efficiency (as has been implemented in Ref.~\cite{act-2010}), we have 
set the CMB spectrum to zero for $\ell>4000$, since the contribution 
at larger multipoles is negligible.
When considering the ACT data, following the earlier work~\cite{act-2010}, 
in the power law case, we have marginalized over the three secondary 
parameters $A_{_{\rm SZ}}$, $A_{_{\rm P}}$ and $A_{_{\rm C}}$, where 
$A_{_{\rm SZ}}$ denotes the Sunyaev- Zeldovich amplitude, $A_{_{\rm P}}$ 
the amplitude for the Poisson power from radio and infrared point sources, 
while $A_{_{\rm C}}$ is the amplitude corresponding to the cluster power.
However, when comparing the oscillatory inflationary potentials with the 
ACT data, we have only marginalized over $A_{_{\rm SZ}}$ and have fixed 
the values of the other two parameters $A_{_{\rm P}}$ and $A_{_{\rm C}}$.

\par 
 
We should mention that we have taken gravitational lensing into account.
Note that, to generate highly accurate lensed CMB spectra, CAMB requires 
$\ell_{\rm max~scalar} \simeq (\ell_{\rm max}+ 500)$, where $\ell_{\rm max}$ 
is, say, the largest multipole moment for which the data is available.
The WMAP seven year data is available up to  $\ell \simeq 1200$, while the 
ACT data is available up to $\ell \simeq 10000$.
For the WMAP seven year data, we set $\ell_{\rm max~scalar}\simeq 1800$, 
and for ACT we choose $\ell_{\rm max~scalar}\simeq 4500$ since we are 
ignoring the data for $\ell > 4000$. 
We set $\ell_{\rm max~tensor}\simeq400$ for all the datasets, as they decay 
down quickly after that. 
ACT has measured only $C_{\ell}^{\rm TT}$, so the constraints from 
polarization, if any, will come only from the WMAP data.

\par

Lastly, since the primordial power spectra that we expect to arise in
the inflationary models of our interest contain repeated patterns 
extending over a wide range of scales, one can expect that equivalent
patterns would be present in the CMB angular power spectrum running over
all angular scales. 
It is well known that the Boltzmann code CAMB uses an effective sampling 
and a highly accurate spline interpolation to determine the CMB angular 
power spectrum over the multipoles of interest~\cite{lewis-2000,camb}. 
However, when the underlying potential power spectra contain oscillations, 
this default technique might not be 
accurate (see Refs.~\cite{pso-1,flauger-2010-2011}; in this context,
also see Ref.~\cite{huang-2012}).
Following a method adopted earlier in a similar context~\cite{pso-1}, 
we incorporate suitable changes in the standard CAMB and CosmoMC packages 
to avoid limited sampling, and evaluate the angular power spectrum at all 
multipoles.

%%%%%%%%%%%%%%%%%%%%%%%%%%%%%%%%%%%%%%%%%%%%%%%%%%%%%%%%%%%%%%%%%%%%%%%%%%%%%%%

\section{Results}

In this section, we shall discuss the results of our analysis.
We shall present the best fit values of the various parameters and also
discuss the resulting primordial and CMB angular power spectra.

%%%%%%%%%%%%%%%%%%%%%%%%%%%%%%%%%%%%%%%%%%%%%%%%%%%%%%%%%%%%%%%%%%%%%%%%%%%%%%%

\subsection{The best fit cosmological and inflationary parameters }

We shall tabulate the best fit parameters in this sub-section. 
We find that our results for the power law case are in good agreement with
the WMAP seven year~\cite{wmap-7} and the ACT results~\cite{act-2010}. 
In fact, we have cross checked our results with and without the tensor 
contribution. 
As stated earlier, we have made use of the three secondary parameters 
$A_{_{\rm SZ}}$, $A_{_{\rm P}}$ and $A_{_{\rm C}}$ when comparing the
power law case with the combined WMAP seven year and ACT data. 
In this case, we obtain the mean value of $A_{_{\rm P}}$ to be 
$16.0$, whereas $A_{_{\rm C}}$ is described by a single tailed 
distribution which suggests that $A_{_{\rm C}} < 8.4$ at $95\%$ 
CL (when the tensors are not taken into account). 
We find that, for the power law spectra, if we fix $A_{_{\rm P}}$ at 
the above-mentioned mean value and set $A_{_{\rm C}}$ to be zero, the 
least squared parameter~$\chi_{\rm eff}^{2}$ changes by a negligible 
amount (in fact, $\Delta\chi_{\rm eff}^{2}\simeq 0.2$--$0.3$), and the 
best fit, the mean values and the deviations too do not change appreciably. 
So, in the case of the two inflationary models of our interest, we have 
set $A_{_{\rm P}}=16.0$, $A_{_{\rm C}}=0$, and have marginalized over
$A_{_{\rm SZ}}$.
In Tab.~\ref{tab:bfv-im} below, we have listed the best fit values 
that we arrive at for the background cosmological parameters and the 
parameters that describe the chaotic inflationary model with superposed 
oscillations and the axion monodromy model. 
%%%%%%%%%%%%%%%%%%%%%%%%%%%%%%%%%%%%%%%%%%%%%%%%%%%%%%%%%%%%%%%%%%%%%%%%%%%%
\begin{table}[!htb]
\label{sine-param}
\begin{center}
\begin{tabular}{|c|c|c|c|}
\hline
& Datasets & {WMAP-$7$}  & {WMAP-$7\,+\,$ACT}\\
\hline
Model & Parameter & Best fit & Best fit\\
\hline 
& $\Omega_{\rm b}\, h^2$ & 0.0220 & 0.0218\\
\cline{2-4} 
& $\Omega_{\rm c}\, h^2$ & 0.1164 & 0.1215\\
\cline{2-4} 
Chaotic & $\theta$ & 1.038 & 1.040\\
\cline{2-4} 
model & $\tau$ & 0.0850 & 0.0876\\ 
\cline{2-4} 
with  & ${\rm ln}\, \l[10^{10}\, m^{2}\r]$ & -0.667 & -0.687\\
\cline{2-4} 
sinusoidal & $\alpha$ & $0.256\times 10^{-3}$ & $0.998\times 10^{-3}$\\
\cline{2-4} 
modulation & $\beta$ & 0.1624 & 0.2106\\
\cline{2-4} 
& $\delta$ & 2.256 & -2.2 \\
\hline
& $\Omega_{\rm b}\, h^2$ & 0.0227 & 0.0223 \\
\cline{2-4} 
& $\Omega_{\rm c}\, h^2$ & 0.1079 & 0.1119\\
\cline{2-4} 
& $\theta$ & 1.040 & 1.041 \\
\cline{2-4} 
Axion & $\tau$ & 0.0921 & 0.0884\\ 
\cline{2-4} 
monodromy & ${\rm ln}\, \l[10^{10}\, \lambda\r]$ & 0.9213 & 0.9332 \\
\cline{2-4} 
model & $\alpha$ & $1.84\times 10^{-4}$ & $1.75\times 10^{-4}$\\
\cline{2-4} 
& $\beta$ & $4.50\times 10^{-4}$ & $5.42\times 10^{-4}$ \\
\cline{2-4} 
& $\delta$ & 0.336 & -0.6342 \\
\hline 
\end{tabular}
\caption{\label{tab:bfv-im} The best fit values for the two inflationary
models on comparing with the WMAP seven year data (denoted as WMAP-$7$
here, and in the following table) alone, and along with the ACT data.}
\end{center}
\end{table}

%%%%%%%%%%%%%%%%%%%%%%%%%%%%%%%%%%%%%%%%%%%%%%%%%%%%%%%%%%%%%%%%%%%%%%%%%%%%%%%

\subsection{The spectra and the improvement in the fit}\label{subsec:chisq}

In Tab.~\ref{tab:chsq}, we have listed the least squares parameter 
$\chi_{\rm eff}^{2}$ for the different models and datasets that we
have considered.
%%%%%%%%%%%%%%%%%%%%%%%%%%%%%%%%%%%%%%%%%%%%%%%%%%%%%%%%%%%%%%%%%%%%%%%%%$$$$$$
\begin{table}[!htb]
\begin{center}
\begin{tabular}{|c|c|c|c|}
\hline
Datasets & WMAP-$7$ & WMAP-$7\,+\,$ACT \\
\cline{1-1}
Model & & \\
\hline
Power law case & 7468.4 & 7500.4 \\
\hline
Chaotic model with & 7468.0 & 7498.2\\
sinusoidal modulation & & \\
\hline
Axion monodromy model & 7455.3 & 7495.2 \\
\hline
\end{tabular}
\caption{\label{tab:chsq}The $\chi_{\rm eff}^{2}$ for the different models 
and datasets that we have considered. 
Note that we have used the Gibbs approach in the WMAP likelihood code to 
calculate the $\chi_{\rm eff}^{2}$ for the CMB $TT$ spectrum at the low 
multipoles (i.e. for $\ell <32$)~\cite{wmap-5,wmap-7}.}
\end{center}
\end{table}
%%%%%%%%%%%%%%%%%%%%%%%%%%%%%%%%%%%%%%%%%%%%%%%%%%%%%%%%%%%%%%%%%%%%%%%%%%%%%%%
From the table it is clear that the monodromy model leads to a much better 
fit with $\chi_{\rm eff}^{2}$ improving by about $13$ in the case of the 
WMAP seven year data and by about $5$ when the ACT data has also been 
included. 
(We shall discuss the reason for this difference in the concluding section.) 
The table also seems to indicate two further points.
Firstly, even though the chaotic model with the sinusoidal modulation does 
not perform as well as the monodromy model, the fact that the model performs 
better when the small scale data from ACT is included suggests that 
oscillations can be favored by the data.
Secondly, oscillations of fixed amplitude in the potential as in the monodromy 
model seem to be more favored by the data than the oscillations of varying 
amplitude as in the case of the chaotic model with sinusoidal modulations.
In fact, this strengthens similar conclusions that has been arrived at 
earlier~\cite{pso-1,pso-2}, wherein Planck scale oscillations of a certain 
amplitude in the primordial spectrum was found to lead to a considerably 
better fit to the data.

\par 

\begin{figure}[!htb]
\subfigure{\includegraphics[width=6cm,angle=-90]{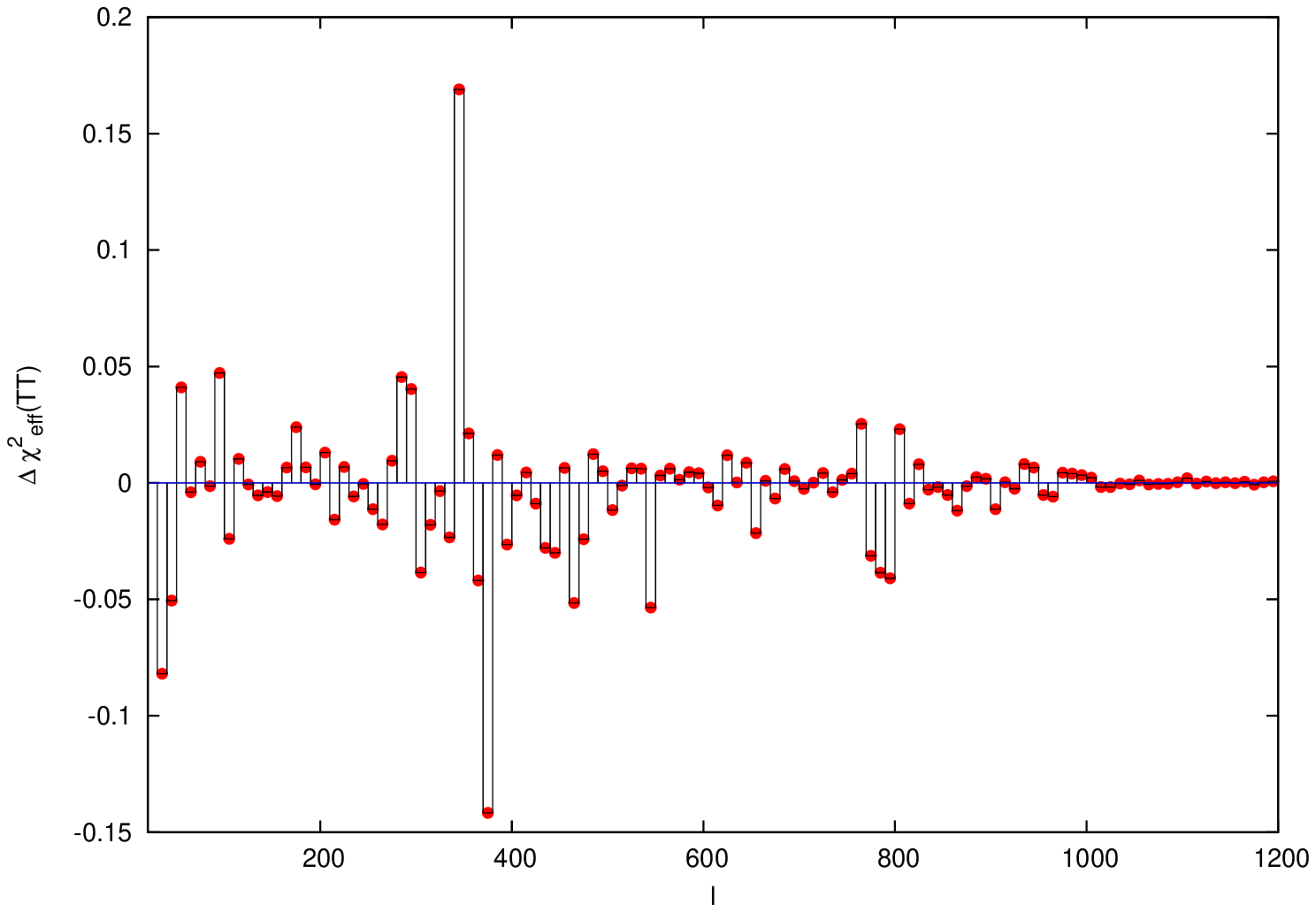}}
\subfigure{\includegraphics[width=6cm,angle=-90]{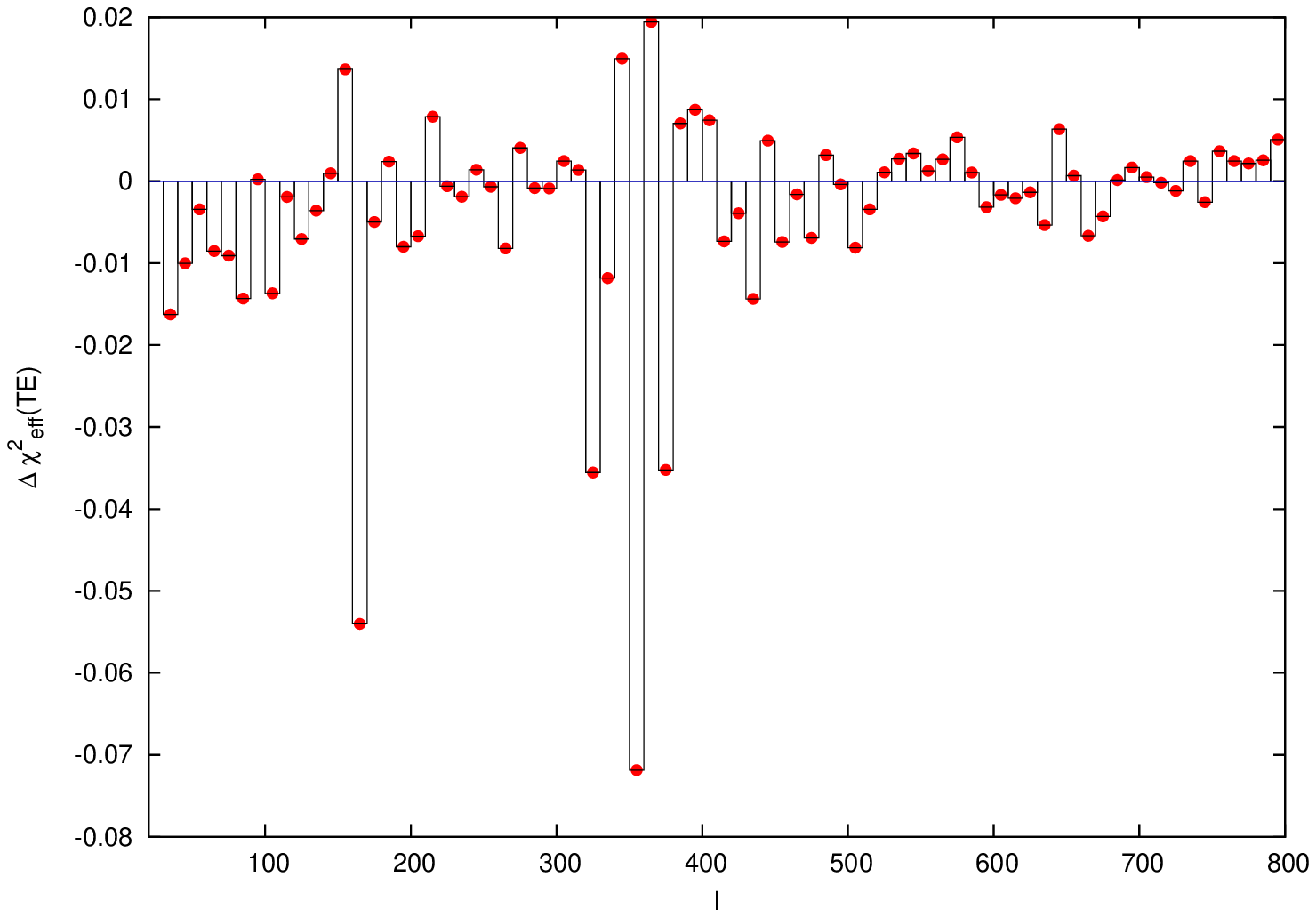}}
\vskip -10 true pt
\caption{\label{fig:deltachsq}The difference in $\chi_{\rm eff}^{2}$
with respect to the reference model, i.e. $\Delta\chi_{\rm eff}^{2}
=[\chi_{\rm eff}^{2}({\rm model})-\chi_{\rm eff}^{2}({\rm power\; law})]$, 
in the case of the axion monodromy model has been plotted as a function 
of the multipole moment for the WMAP seven year data, after binning in 
the multipole space with $\ell_{\mathrm{bin}}=10$.
While the figure on top corresponds to the WMAP seven year $TT$ data
(for $\ell>32$), the lower one is for the $TE$ data (for $\ell>24$).}
\end{figure}

\begin{figure}[!htb]
\begin{center}
\hskip 25pt
\resizebox{240pt}{160pt}{\includegraphics{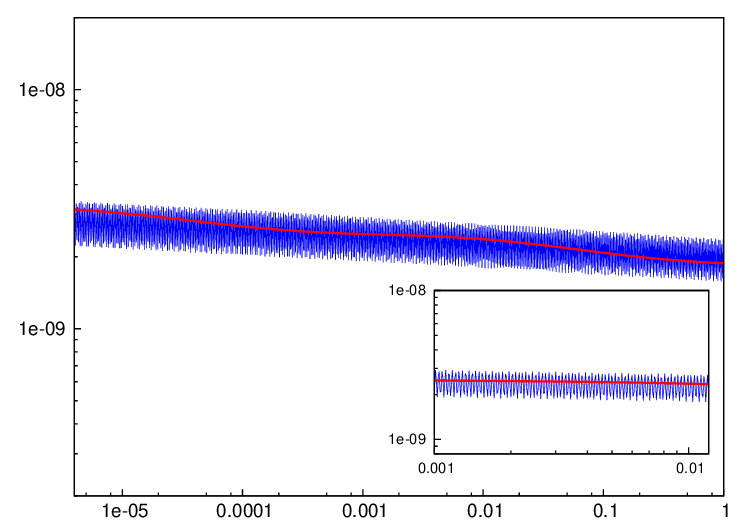}}
\vskip -100 true pt 
\hskip -230 true pt ${\cal P}_{_{\rm S}}(k)$
\vskip 92 pt
\hskip 20 true pt $k$ 
\caption{\label{fig:sps}The scalar power spectra corresponding to the 
best fit values of the WMAP seven year data for the two inflationary
models that we have considered.
The solid red and the solid blue lines describe the scalar power spectra 
in the cases of the chaotic model with a sinusoidal modulation and the 
axion monodromy model, respectively.
The spectrum corresponding to the best fit power law model would 
essentially be the same as in the chaotic model with sinusoidal 
modulations, but without any oscillations.
The inset highlights the extraordinary extent of persistent oscillations 
in the case of the monodromy model.}
\end{center}
\end{figure}

\begin{figure}[!htb]
\begin{center}
\hskip 30pt
% \resizebox{230pt}{170pt}[angle=-90]{\includegraphics{cl.ps}}
\subfigure{\includegraphics[width=6cm,angle=-90]{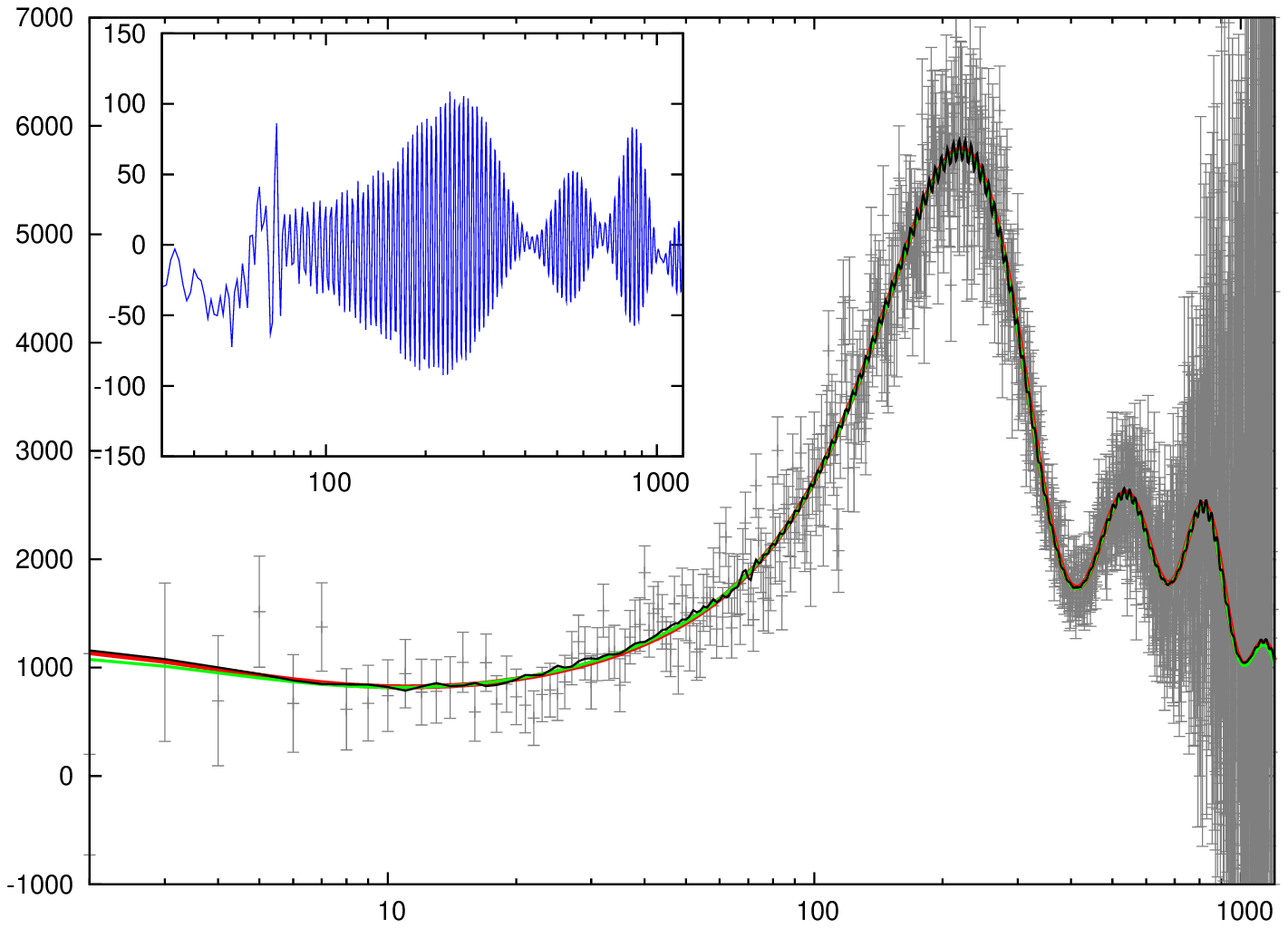}}
% \subfigure{\includegraphics[width=4cm,angle=-90]{cl-diff.ps}}
\vskip -145 true pt 
\hskip -240 true pt 
\rotatebox{90}{$\l[\ell\, (\ell+1)\; 
C_{\ell}^{TT}/(2\,\pi)\r]\; \mu\, {\rm K}^{2}$}
\vskip 30 true pt 
\hskip 12 true pt $\ell$
\caption{\label{fig:cltt}The CMB $TT$ angular power spectra corresponding 
to the best fit values of the different models for the WMAP seven year 
data.
The solid red, solid green and the black curves correspond to the 
power law model, the chaotic model with sinusoidal modulation and the 
axion monodromy model, respectively.
The gray circles with error bars denote the WMAP seven year {\it unbinned}\/ 
data.
The inset highlights the difference in the angular power spectrum between the
monodromy model and the power law case.
In the case of the axion monodromy model, the tiny and continued oscillations 
in the power spectrum lead to small improvements in the fit to the data over 
a wide range of multipoles, which eventually add up to a good extent.}
\end{center}
\end{figure}

\begin{figure}[!htb]
\begin{center}
\hskip 25pt
% \resizebox{230pt}{170pt}{\includegraphics{abs_te_ee.eps}}
\subfigure{\includegraphics[width=6cm,angle=-90]{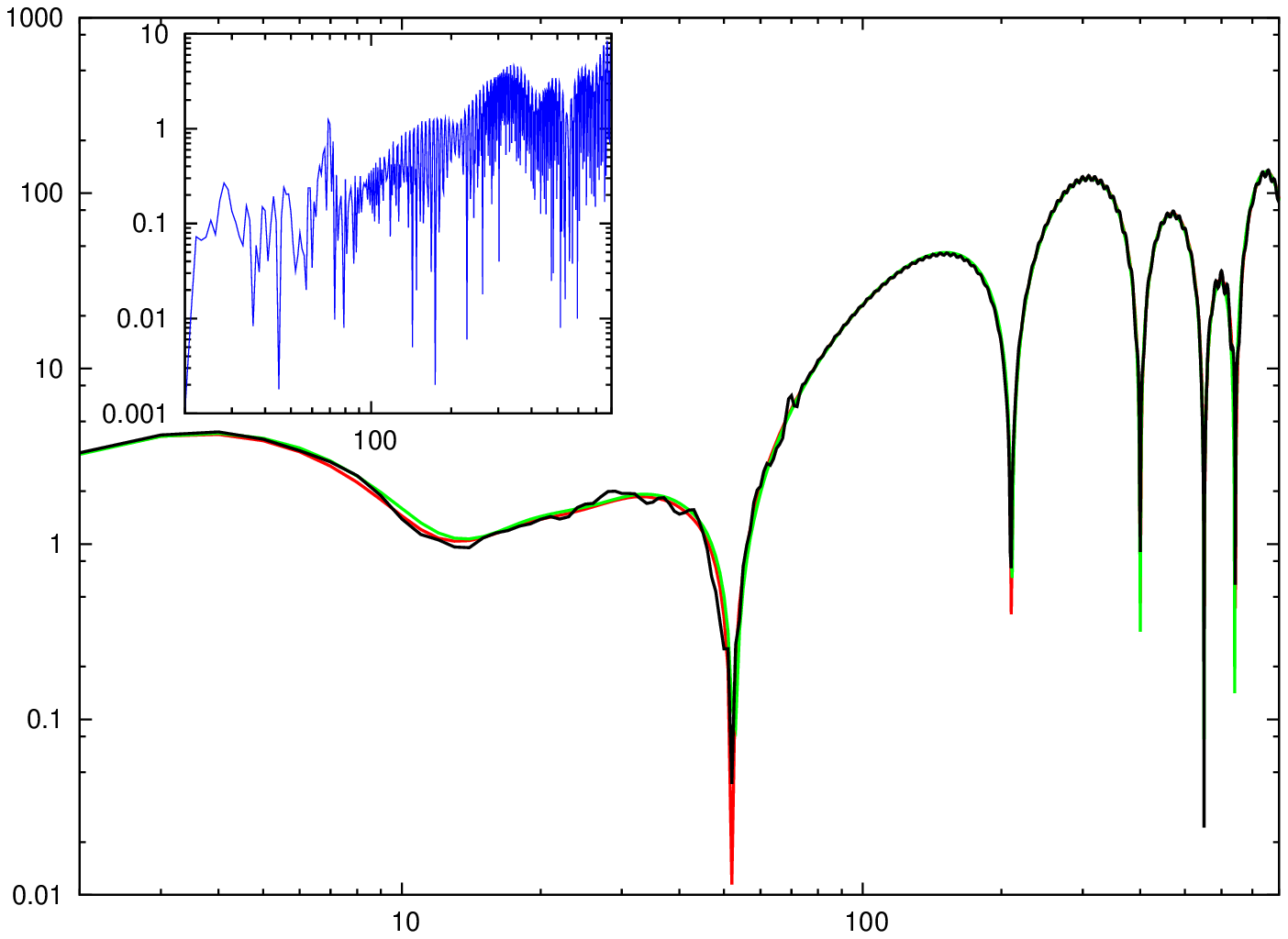}}
\subfigure{\includegraphics[width=6cm,angle=-90]{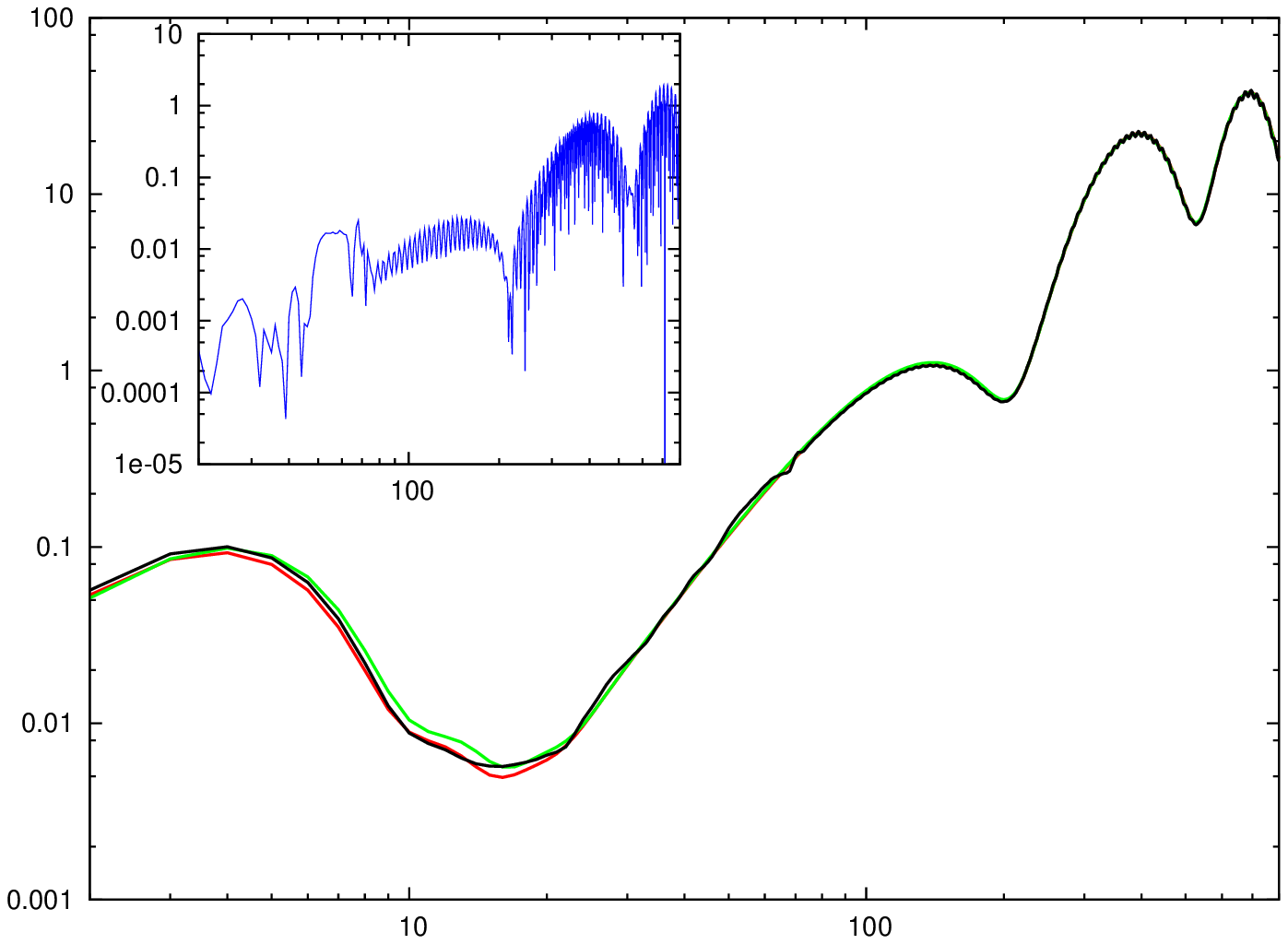}}
% \subfigure{\includegraphics[width=4cm,angle=-90]{ee-diff.ps}}
\vskip -330 true pt \hskip -250 true pt 
\rotatebox{90}{$\l[\ell\, (\ell+1)\;
C_{\ell}^{TE}/(2\,\pi)\r]\; \mu\, {\rm K}^{2}$}
\vskip 75 true pt \hskip -230 true pt 
\rotatebox{90}{$\l[\ell\, (\ell+1)\;
C_{\ell}^{EE}/(2\,\pi)\r]\; \mu\, {\rm K}^{2}$}
\vskip 30 true pt \hskip 20 true pt $\ell$
\caption{\label{fig:clteee} The CMB $TE$ and the $EE$ angular power spectra 
corresponding to the best fit values of the different models for the WMAP 
seven year data.
The solid red, green and black curves represent the $TE/EE$ spectrum (in fact, 
magnitude of TE spectrum) in the power law case, the chaotic model with sinusoidal 
modulations and the axion monodromy model, respectively.
As in the earlier figure, the insets highlight the difference in the $TE/EE$ 
spectrum between the monodromy model and the power law case.}
\end{center}
\end{figure}

It is now interesting to enquire as to whether there exist localized windows
of multipoles over which the improvement in the fit occurs.
We find that, in the case of the chaotic model with sinusoidal modulations,
as far as the WMAP seven year data is concerned, there is an improvement of 
at most unity in all the multipoles combined.
For the monodromy model, the improvement at the low multipoles (i.e. for 
$\ell < 32$) is just about $3$ in the WMAP seven year $TT$ data, and there 
is hardly any improvement in the fit from the available $TE$ and $EE$ data.
We find that most of the improvement occurs at the higher multipoles in the
$TT$ data.
In Fig.~\ref{fig:deltachsq}, after binning suitably, we have plotted 
the difference $\Delta\chi_{\rm eff}^{2}=[\chi_{\rm eff}^{2}({\rm model})
-\chi_{\rm eff}^{2}({\rm power\; law})]$, as a function of the multipoles 
for the WMAP seven year $TT$ and $TE$ data in the case of the axion 
monodromy model.
%%%%%%%%%%%%%%%%%%%%%%%%%%%%%%%%%%%%%%%%%%%%%%%%%%%%%%%%%%%%%%%%%%%%%%%%%%%%%%%
%%%%%%%%%%%%%%%%%%%%%%%%%%%%%%%%%%%%%%%%%%%%%%%%%%%%%%%%%%%%%%%%%%%%%%%%%%%%%%%
It is clear from the figure that the source of the improvement in the fit
is not confined to any specific set of multipoles, and it arises 
due to small increments that accrue over the entire range of available data.
In Figs.~\ref{fig:sps} and~\ref{fig:cltt}, we have plotted the scalar power
spectra and the corresponding CMB $TT$ angular power spectra for the best
fit values of the WMAP seven year data in the two inflationary models that
we have considered.
%%%%%%%%%%%%%%%%%%%%%%%%%%%%%%%%%%%%%%%%%%%%%%%%%%%%%%%%%%%%%%%%%%%%%%%%%%%%%%%
%%%%%%%%%%%%%%%%%%%%%%%%%%%%%%%%%%%%%%%%%%%%%%%%%%%%%%%%%%%%%%%%%%%%%%%%%%%%%%%  
%%%%%%%%%%%%%%%%%%%%%%%%%%%%%%%%%%%%%%%%%%%%%%%%%%%%%%%%%%%%%%%%%%%%%%%%%%%%%%%
And, in Fig.~\ref{fig:clteee}, we have plotted the corresponding CMB $EE$ 
angular power spectra and $TE$ amplitude for all the models, including the
power law case.
%%%%%%%%%%%%%%%%%%%%%%%%%%%%%%%%%%%%%%%%%%%%%%%%%%%%%%%%%%%%%%%%%%%%%%%%%%%%%%%

%%%%%%%%%%%%%%%%%%%%%%%%%%%%%%%%%%%%%%%%%%%%%%%%%%%%%%%%%%%%%%%%%%%%%%%%%%%%%%%

\section{Can Planck see the oscillations?}

In this section, we shall discuss the extent to which the data from 
Planck---that is expected to be coming forth in the very near 
future---will be able to constrain the presence and characteristics 
of extended features in the primordial spectrum.

\par 

Many of the parameters of inflationary models of the type that we are
considering here can have a credible physical influence on the cosmological 
data, even if their presence has not yet been detected. 
It is expected that data from current missions such as Planck and beyond 
would be able to determine many of the presently unknown effects of the 
cosmological parameters. 
When performing a parameter error forecast for future observations, it is
the Fisher matrix formalism that is commonly adopted. 
The error bars on the additional parameters are estimated from the derivatives 
of the observables with respect to the model parameters around the best fit 
point (for a discussion, see, for example, Ref.~\cite{dodelson-2003}).
Such an analysis assumes that the likelihood of the cosmological parameters
approximates a Gaussian multivariate. 
However, parameter degeneracies can occur where certain combinations of 
the parameters are not well constrained by the data. 
Also, the probability distribution of the parameters defined over a finite 
range may occasionally fail to converge at the boundaries. 
These lead to considerable deviations from the assumption of a multivariate 
Gaussian function.

\par

We arrive at the possible constraints on the parameters using a different 
technique wherein we make suitable modifications to the CosmoMC code with a 
publicly available add-on code FuturCMB~\cite{futurecmb,perotto-2006,pahud-2006}.
We firstly generate a simulated dataset for Planck, using realistic isotropic 
noise levels and the sky coverage fraction $f_{\mathrm{sky}}$. 
We consider only isotropic noise modeled as spatially uniform Gaussian white 
noise. 
This ensures that the noise term is diagonal in the multipole space.
The CMB angular power spectrum generated from the best fit parameters of the 
axion monodromy model (cf.~Tab.~\ref{tab:bfv-im}) using the WMAP seven year 
data is treated as the fiducial power spectrum for generating the Planck mock 
data. 
We use this simulated data and incorporate the `all\_l\_exact' data format in
the CosmoMC code~\cite{all-ell} to extract the projected parameter errors by 
sampling the likelihood and estimating the marginalized probability distribution 
in the parameter space. 
As we mentioned above, it is expected that this procedure would be more reliable 
than the Fisher matrix analysis since there is no assumption on the likelihood 
functions of these parameters being multivariate Gaussian distributions. 

\par

\begin{figure}[!htb]
\begin{center}
\hskip -10pt
\resizebox{250pt}{225pt}{\includegraphics{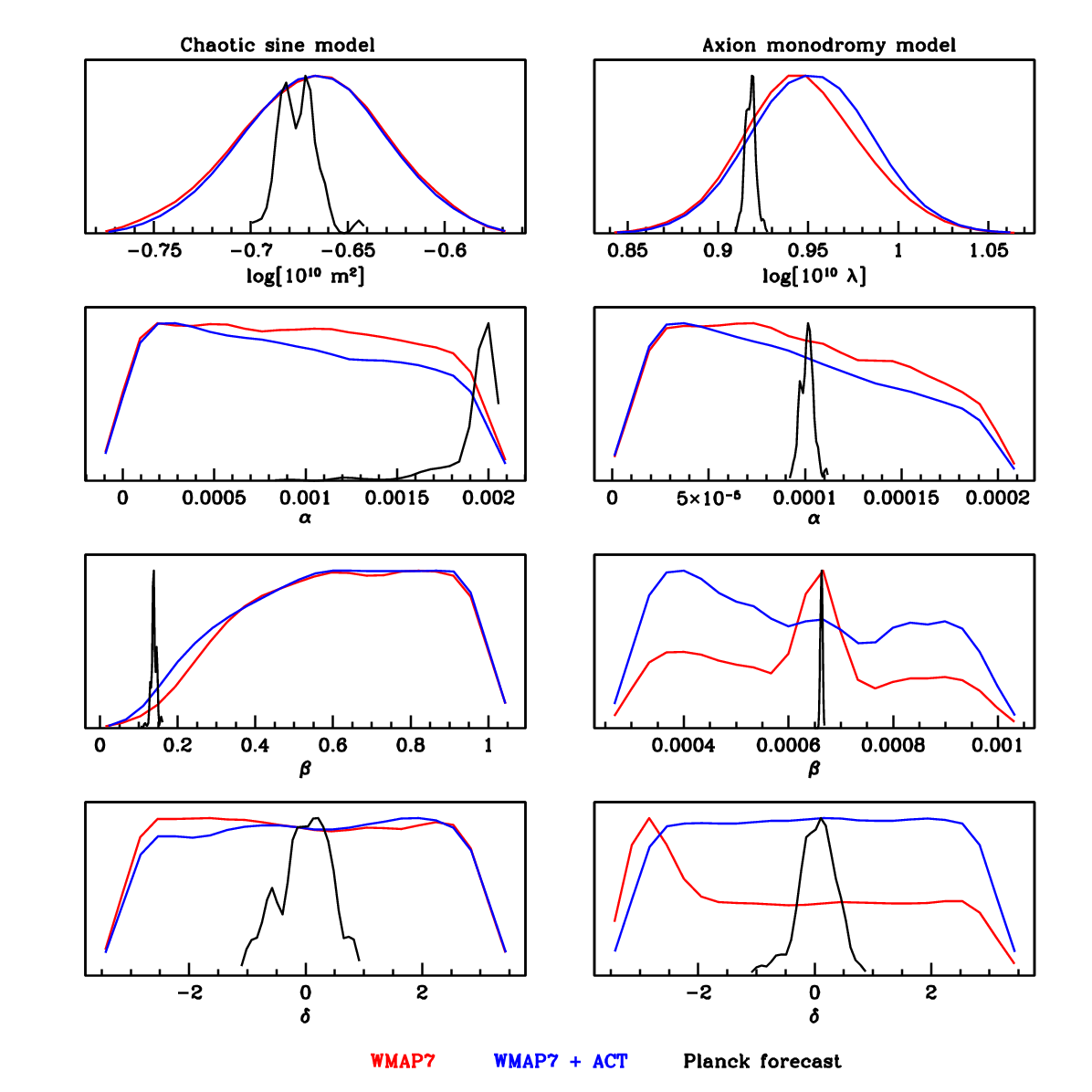}}
\end{center}
\caption{\label{fig:one-d}One dimensional distributions of the inflationary 
model parameters from the WMAP-$7$, WMAP-$7$ $+$ ACT and the Planck simulated 
data.
We have plotted the constraints on the parameters $m$, $\alpha$, $\beta$ and 
$\delta$ of the chaotic model with sinusoidal modulations (on the left column) 
and the parameters $\lambda$, $\alpha$, $\beta$ and $\delta$ for the axion 
monodromy model (on the right column).
It is evident that the simulated Planck data tightens the bounds on the 
parameters substantially.}
\end{figure}

We have plotted the resulting one dimensional distributions for the 
inflationary parameters in Fig.~\ref{fig:one-d}.
The figure contains the constraints from the WMAP-7, WMAP-7 $+$ ACT as well
as the Planck simulated data for the original parameter $m$ of the chaotic
model and the amplitude, frequency and phase parameters $\alpha$, $\beta$ 
and $\delta$ of the superimposed sinusoidal modulations.
For the axion monodromy model, we have plotted the distributions for the 
initial parameter $\lambda$ and the parameters $\alpha$, $\beta$ and $\delta$ 
characterizing the oscillations for the same datasets. 
The one dimensional distribution for $\delta$ is very flat suggesting the lack 
of a specific region in the parameter space of $\delta$ which provides a good 
fit to the data. 
However, $\delta$ is a phase parameter in the oscillating inflaton potential 
which shifts the oscillations within one period in our analysis; a flat 
likelihood does not necessarily indicate the failure of the particular model. 
%%%%%%%%%%%%%%%%%%%%%%%%%%%%%%%%%%%%%%%%%%%%%%%%%%%%%%%%%%%%%%%%%%%%%%%%%%%%%%%
%%%%%%%%%%%%%%%%%%%%%%%%%%%%%%%%%%%%%%%%%%%%%%%%%%%%%%%%%%%%%%%%%%%%%%%%%%%%%%%

\par

\begin{figure}[!htb]
\begin{center}
\resizebox{225pt}{175pt}{\includegraphics{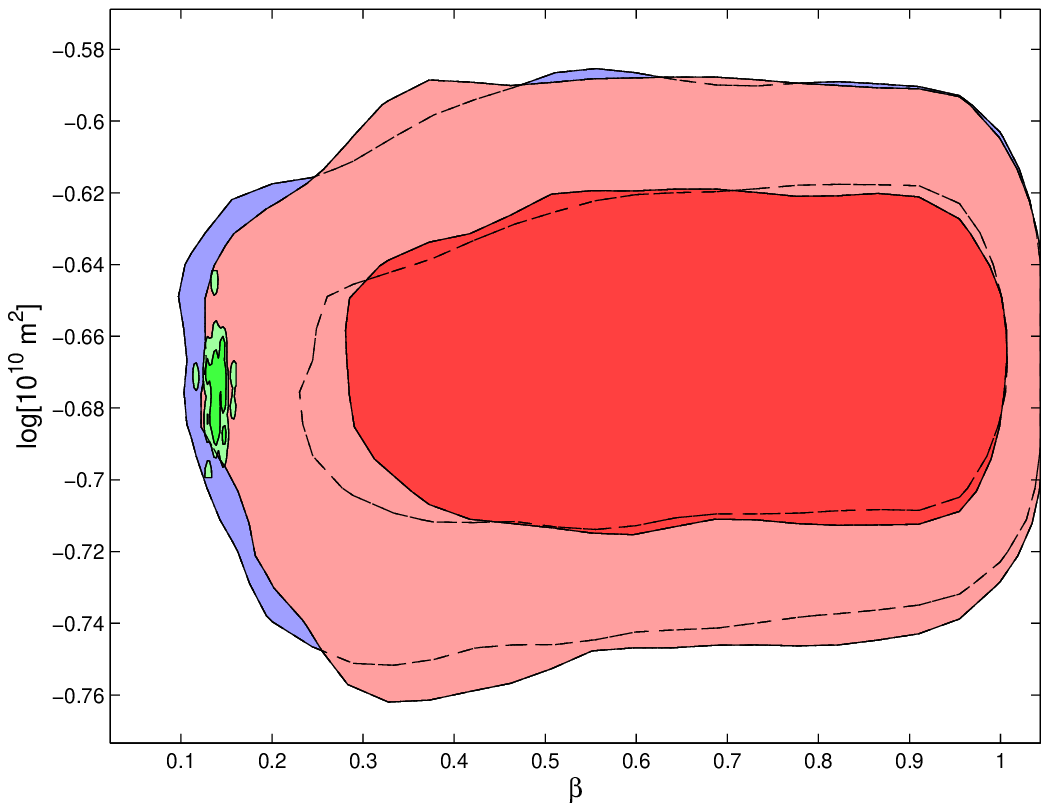}}
\vskip 10 pt
\resizebox{225pt}{175pt}{\includegraphics{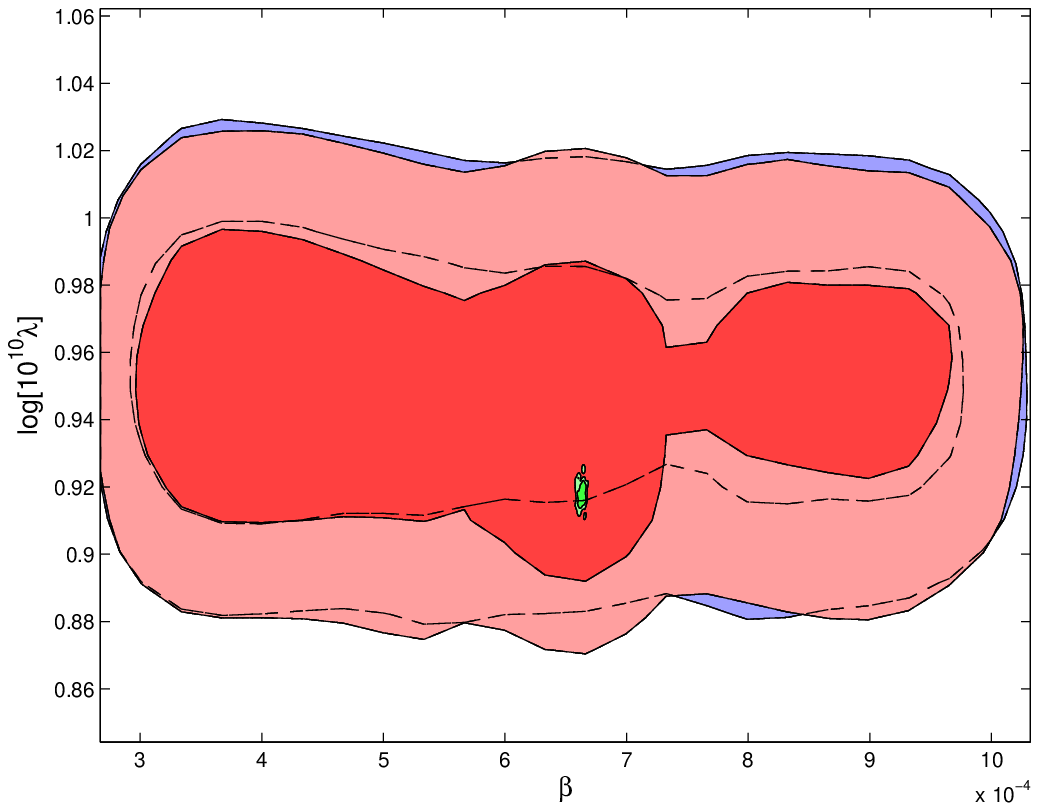}}
\end{center}
\caption{\label{fig:two-d}The joint two dimensional constraints on the 
inflationary model parameters from the WMAP-$7$, WMAP-$7$ + ACT and the 
Planck simulated data.
The figure on top illustrates the joint constrains on the parameters $m$ 
and $\beta$ that characterize the chaotic model with sinusoidal modulations,
while the lower figure displays the joint constraints on the parameters 
$\lambda$ and $\beta$ that describe the axion monodromy model.
Note that the red contours are the $1$-$\sigma$ and the $2$-$\sigma$ 
constraints from WMAP-$7$, the blue contours are from WMAP-$7$ + ACT, 
while the green contours are from the simulated Planck data. 
(The broken black lines on the red contours are traces of the underlying 
blue contours from the WMAP-$7$ + ACT data.)
It is again clear that the simulated Planck data constrains the parameters
considerably more than the available data.}
\end{figure}

Fig.~\ref{fig:two-d} contains the two dimensional contour plots for the 
parameters $m$ and $\beta$ for the sinusoidal model and $\lambda$ and 
$\beta$ for the monodromy model. 
Again, we have displayed the constraints from all the three datasets.  
%%%%%%%%%%%%%%%%%%%%%%%%%%%%%%%%%%%%%%%%%%%%%%%%%%%%%%%%%%%%%%%%%%%%%%%%%%%%%%%
%%%%%%%%%%%%%%%%%%%%%%%%%%%%%%%%%%%%%%%%%%%%%%%%%%%%%%%%%%%%%%%%%%%%%%%%%%%%%%%
It can be easily perceived from these set of figures that the Planck
data leads to much tighter bounds on the inflationary parameters than
the currently available data for the same range of priors for the
various parameters.

%%%%%%%%%%%%%%%%%%%%%%%%%%%%%%%%%%%%%%%%%%%%%%%%%%%%%%%%%%%%%%%%%%%%%%%%%%%%%%%

\section{Discussion}

In this work, our main aim has been to investigate if the CMB data support 
certain non-local features---i.e. a certain repeated and characteristic 
pattern that extends over a wide range of scales---in the primordial scalar 
power spectrum.
With this goal in mind, we have studied two models of inflation, both of 
which contain oscillatory terms in the inflaton potential.
The oscillations in the potential produces oscillations in the slow roll 
parameters, which in turn generate oscillations in the primordial as well 
as the CMB power spectra. 
Earlier work in this context had utilized the analytical expressions for the 
primordial power spectra, obtained in the slow roll approximation, to compare 
such models with the data~\cite{pahud-2009,flauger-2010-2011,kobayashi-2011}.
Instead, we have used an accurate and efficient numerical code to arrive at
the inflationary scalar and tensor power spectra.
In fact, in order to ensure a good level of accuracy, rather than evolve 
a finite set of modes and interpolate, we have evolved and computed the 
inflationary perturbation spectra for all the modes that is required by 
CAMB to arrive at the corresponding CMB angular power spectra.  
While this reflects the extent of the numerical accuracy of our computations, 
the efficiency of the code can be gauged by the fact that we have able to been
able to complete the required runs within a reasonable amount of time despite 
such additional demands. 

\par

Prior experience, gained in a different context, had already suggested 
the possibility that small and continued oscillations in the scalar 
power spectra can lead to a better fit to the data~\cite{pso-1}.
This experience has been corroborated by the 
earlier~\cite{pahud-2009,flauger-2010-2011,kobayashi-2011} and our current 
analysis (in this context, see, however, Ref.~\cite{huang-2012}; we 
shall comment further on this point below).
We find that, oscillations, such as those occur in the axion monodromy model
lead to a superior fit to the data. 
In fact, as far as the WMAP seven year data is concerned, on evaluating the 
CMB angular power spectrum at all the required multipoles without any 
interpolation, we obtain an improvement of about $13$ in the least squared 
parameter $\chi_{\rm eff}^{2}$ for the axion monodromy model, just as the 
earlier analytical efforts had (see the first of the two references
in Refs.~\cite{flauger-2010-2011}).
The time taken to compute the uninterpolated inflationary power spectra 
depends not only on the number of points required, but also on the 
frequency of the oscillations in the inflaton potentials that we have 
considered.
In the case of the axion monodromy model, over the range of parameters 
that we have worked with, our code takes about $3$-$12$ seconds to 
calculate the inflationary power spectra (both scalar and tensor) for the 
nearly $2000$ $k$-points which are required by CAMB. 
While such a level efficiency seems adequate for comparing the models of
our interest with the WMAP seven year data, we found that evaluating the
uninterpolated CMB angular power spectra for comparing with the WMAP as
well as the ACT data sets did not prove to be feasible in reasonable 
amount of time. 
As a result, we were forced to use the default, interpolated 
CMB angular power spectra obtained by CAMB in this situation. 
It is for this reason that we have not been able to achieve an equivalent
improvement in the $\chi_{\rm eff}^{2}$ for the monodromy model when the
ACT data has been included.

\par

Nevertheless, we believe that the limited level of comparison with the ACT 
data has its own role to play.
The ACT data we have used in our analysis is the binned data provided in the 
ACT likelihood software. 
For a small sky coverage experiment such as ACT, a lot of systematics are
involved in reconstructing the unbinned data. 
The difference in $\chi_{\rm eff}^2$ values using only the WMAP dataset and 
both the WMAP and ACT datasets approximately corresponds to the number of 
binned data points in the ACT dataset. 
The reason we have incorporated the ACT dataset is to cover the large multipole 
regime in the angular power spectrum. 
For the monodromy model, we see that the tiny oscillations do continue till 
small scales which does not overlap with the WMAP seven year dataset, but 
can be probed using the ACT dataset. 
Combining the two datasets, one can form an informed estimate of the model 
parameters over a wide range of angular scales.

\par 

In addition to comparing with the already available data, we have also 
discussed on the extent to which Planck may be able to constrain the 
parameters that describe the oscillatory terms in the potential.
Rather than adopt the standard method of forecasting for the model 
parameters using the Fisher matrix, we have been able to arrive at
the constraints with suitable modifications to CosmoMC.
We believe that the method we have adopted is more reliable than the Fisher
matrix approach which does not work equally well when the parameters are not 
described by multivariate Gaussian distributions. 
The one-dimensional marginalized distributions and the two-dimensional
contours for the parameters of the inflationary models that we have 
arrived at show that future full sky CMB data sets such as Planck would 
be capable of narrowing the constraints on these parameters considerably.

\par

Finally, before closing, it is important that we 
comment on a recent work wherein it has been argued that fine
features in the primordial spectrum as generated by models such 
as the axion monodromy model have been not conclusively detected
by the data~\cite{huang-2012}.
It should be emphasized that, in this work, we have evaluated an 
uninterpolated CMB angular power spectrum while comparing the 
models with the data.
Moreover, the resulting best fit CMB angular power spectra 
(cf. Figs.~\ref{fig:cltt} and~\ref{fig:clteee}) do indeed contain 
the tiny and persistent features encountered in the recent (see 
Fig.~4 of Ref.~\cite{huang-2012}) as well as the earlier 
work~\cite{pso-1,pso-2,flauger-2010-2011}.
Also, as we have highlighted before, the results from our numerical 
evaluation of the inflationary power spectra largely match the earlier 
results arrived at from the corresponding analytical spectra.
While it may be true that the evidence for the oscillations may still
not be conclusive, repeated analyses have unambiguously pointed to the
fact that they are more favored by the data than a simpler and smooth
primordial spectrum.
As we have argued, we believe that Planck may be able to provide 
conclusive evidence in this regard.

%%%%%%%%%%%%%%%%%%%%%%%%%%%%%%%%%%%%%%%%%%%%%%%%%%%%%%%%%%%%%%%%%%%%%%%%%%%%%%%

\section*{Acknowledgments}

We acknowledge the use of the high performance computing facilities at 
the Harish-Chandra Research Institute, Allahabad, India, as well as at 
the Inter-University Centre for Astronomy and Astrophysics, Pune, India.
TS and MA acknowledge support from the Swarnajayanti Fellowship, 
Department of Science and Technology, India.
DKH would like to thank Sanjoy Biswas for useful discussions on numerical 
procedures.
We would also like to acknowledge the use of the CosmoMC 
package~\cite{cosmomc}, and the data products provided by the WMAP science 
team~\cite{lambda}, the ACT and the Planck missions.

%%%%%%%%%%%%%%%%%%%%%%%%%%%%%%%%%%%%%%%%%%%%%%%%%%%%%%%%%%%%%%%%%%%%%%%%%%%%%%%

%%%%%%%%%%%%%%%%%%%%%%%%%%%%%%%%%%%%%%%%%%%%%%%%%%%%%%%%%%%%%%%%%%%%%%%%%%%%%%%
\end{document}